\documentstyle[prc,aps,epsfig,multicol]{revtex}
\tightenlines

\begin{document}
\title{Charmonium absorption cross section by nucleon}
\bigskip
\author{W. Liu, C. M. Ko, and Z. W. Lin}
\address{Cyclotron Institute and Physics Department, Texas A\&M University,
College Station, Texas 77843-3366}
\maketitle

\begin{abstract}
The $J/\psi$ absorption cross section by nucleon is studied using
a gauged SU(4) hadronic Lagrangian but with empirical particle masses,
which has been used previously to study the $J/\psi$ absorption 
cross section by pion and rho meson. Including both two-body
and three-body final states, we find that with a cutoff parameter 
of 1 GeV at interaction vertices involving charm hadrons, 
the $J/\psi-N$ absorption is at most 5 mb and is consistent with 
that extracted from $J/\psi$ production from both photo-nuclear 
and proton-nucleus reactions.

\end{abstract}
 
\begin{multicols}{2}

\section{introduction}

Two main mechanisms for $J/\psi$ suppression observed
in relativistic heavy ion collisions \cite{jpsi} are the 
dissociation by the quark-gluon plasma \cite{matsui} and 
the absorption by comoving hadrons, mainly pions and rho 
mesons \cite{comover}. The cross sections of $J/\psi$ absorption
by hadrons are, unfortunately, not well determined. In the
perturbative QCD approach \cite{kharzeev}, based on the dissociation of
charmonium bound states by energetic gluons inside hadrons, 
the dissociation cross section increases monotonously with 
the kinetic energy of hadrons and has a value of only about $0.1$ mb 
at 0.8 GeV. On the other hand, both the quark-interchange
model \cite{qem} based on the gluon-exchange potential and the meson-exchange 
model \cite{haglin,lin,oh} based on hadronic Lagrangians
give $J/\psi$ absorption cross sections by pion and rho meson
which are more than an order of magnitude larger, i.e., a few mb.
A similar magnitude for the $J/\psi-\pi$ absorption cross section
has also been obtained in the QCD sum rules \cite{nielsen}. 

Since the $J/\psi$ absorption cross sections by pion and rho meson
cannot be directly measured, it is useful to find empirical information 
which can constrain their values. One such constraint is the $J/\psi$ 
absorption cross section by nucleon, as this process can be viewed as 
$J/\psi$ absorption by the virtual pion and rho meson cloud of the nucleon.
From $J/\psi$ production in photo-nucleus reactions, the cross section 
of $J/\psi$ absorption by nucleon can be extracted, and its magnitude 
has been found to be about 4 mb \cite{photo}. The $J/\psi-N$ absorption 
cross section has also been extracted from proton-nucleus collisions at proton
energies from 200 to 800 GeV, and the empirical value is about 7 mb
\cite{pa}. 

In the meson-exchange model of Refs. \cite{haglin,lin,oh}, the 
interaction Lagrangians between pseudoscalar and vector mesons are 
obtained from the SU(4) invariant free Lagrangian for pseudoscalar 
mesons by treating vector mesons as gauge particles. This then leads
to not only pseudoscalar-pseudoscalar-vector-meson couplings but 
also three-vector-meson and four-point couplings. Since the SU(4)
symmetry is explicitly broken by hadron masses, empirical hadron masses 
are used in the Lagrangian. Furthermore, values for the coupling constants 
are taken either from empirical information if they are available or
from theoretical models, such as the vector meson dominance model and
the QCD sum rules. Otherwise, they are determined by using relations
derived from the SU(4) symmetry. In this paper, we shall generalize
this Lagrangian to study the $J/\psi$ absorption cross section
by nucleon and to see if its magnitude is consistent with that
extracted from $J/\psi$ production in photo-nucleus and proton-nucleus
reactions.

This paper is organized as follows. In Section \ref{pion}, we first 
consider $J/\psi$ absorption by nucleon via pion and rho meson 
exchange. The process of $J/\psi$ absorption by nucleon via charm 
exchange is studied in Section \ref{charm}. The effect due to the 
anomalous parity interaction of $J/\psi$
with charm mesons is studied in Section \ref{anomalous}.
In Section \ref{total}, the total $J/\psi$ absorption cross section
by nucleon is given. Finally, conclusions and discussions are given 
in Section \ref{conclusion}. An Appendix is included to derive
the SU(4) relations for some of the coupling constants involving
charm hadrons.

\section{$J/\psi$ absorption by nucleon via pion and rho meson exchange}
\label{pion} 

\begin{figure}[h]
\centerline{\epsfig{file=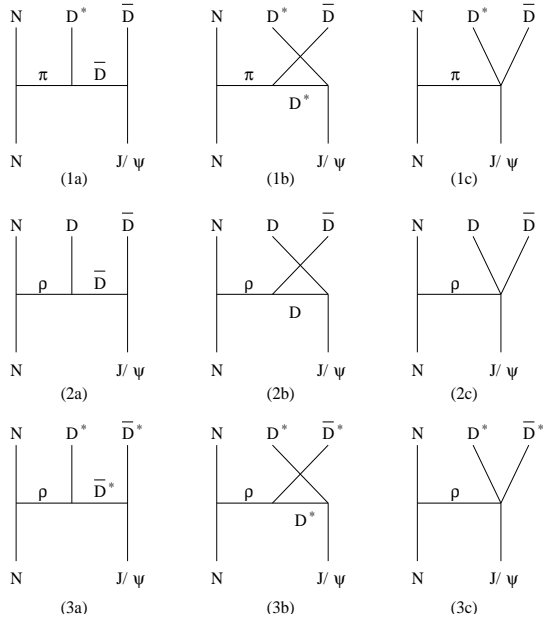,width=2.8in,height=3.2in,angle=0}}
\vspace{0.5cm}
\caption{$J/\Psi$ absorption by nucleon via pion and rho meson exchanges.}
\label{diagram1}
\end{figure}

Possible processes for $J/\psi$ absorption by nucleon involving
its virtual pion and rho meson cloud are $J/\psi N\to D^*\bar DN 
(\bar D^*DN)$, $J/\psi N\to D\bar DN$, and $J/\psi N\to D^*\bar D^*N$, 
as shown by the diagrams in Fig. \ref{diagram1}. The cross sections 
for these processes can be evaluated using the Lagrangians introduced 
in \cite{haglin,lin,oh} for $J/\psi$ absorption by real pion and
rho meson and in Ref. \cite{lin2} for charm meson scattering
by these hadrons. 

The interaction Lagrangian densities that are relevant to the
present study are given as follows:
\begin{eqnarray} 
{\cal L}_{\pi NN} & = & -ig_{\pi NN}\bar{N}\gamma_{5}\vec{\tau}N\cdot\vec{\pi},
\label{pnn}\\ 
{\cal L}_{\rho NN} & = & g_{\rho NN}\bar{N}(\gamma^{\mu}\vec{\tau}\cdot
\vec{\rho}_{\mu}+\frac{\kappa_{\rho}}{2m_{N}}\sigma^{\mu\nu}\vec{\tau}
\cdot\partial_{\mu}\vec{\rho}_{\nu})N,\label{rnn}\\
{\cal L}_{\pi DD^{*}} & = & ig_{\pi DD^{*}}D^{*\mu}\vec{\tau}\cdot(\bar{D}
\partial_{\mu}\vec{\pi}-\partial_{\mu}\bar{D}\vec{\pi})+{\rm H.c.}, \\
{\cal L}_{\rho DD} & = & ig_{\rho DD}(D\vec{\tau}\partial_{\mu}\bar{D}-
\partial_{\mu}D\vec{\tau}\bar{D})\cdot\vec{\rho}^{\mu},\\  
{\cal L}_{\rho D^{*}D^{*}} & = & ig_{\rho D^{*}D^{*}}[(\partial_{\mu}D^{*\nu}
\vec{\tau}\bar{D}^{*}_{\nu}-D^{*\nu}\vec{\tau}\partial_{\mu}
\bar{D}^{*}_{\nu})\cdot\vec{\rho}^{\mu} \nonumber\\
&+&(D^{*\nu}\vec{\tau}\cdot\partial_{\mu}\vec{\rho}^{\nu}
-\partial_{\mu}D^{*\nu}\vec{\tau}\cdot\vec{\rho}_{\nu})\bar{D}^{*\mu}
\nonumber\\      
&+&D^{*\mu}(\vec{\tau}\cdot\vec{\rho}^{\nu}\partial_{\mu}
\bar{D}^{*}_{\nu}-\vec{\tau}\cdot\partial_{\mu}\vec{\rho}^{\nu}
\bar{D}^{*}_{\nu})],\\
{\cal L}_{\psi DD} & = & ig_{\psi DD}\psi^{\mu}[D\partial_\mu\bar D)
-(\partial_{\mu}D)\bar D],\\
{\cal L}_{\psi D^{*}D^{*}} & = & ig_{\psi D^{*}D^{*}}[\psi^{\mu}
(\partial_{\mu}D^{*\nu}\bar{D}^{*}_{\nu}-D^{*\nu}\partial_{\mu}
\bar{D}^{*}_{\nu}) \nonumber\\ 
&+&(\partial_{\mu}\psi ^\nu D^{*}_{\nu}-\psi^{\nu}
\partial_{\mu}D^{*}_{\nu})\bar{D}^{*\mu}\nonumber\\
&+&D^{*\mu}(\psi^{\nu}
\partial_{\mu}D^{*}_{\nu}-\partial_{\mu}\psi^{\nu}\bar{D}^{*}_{\nu}), \\ 
{\cal L}_{\pi \psi DD^{*}} & = & -g_{\pi \psi DD^{*}}\psi^{\mu}(D^{*}_{\mu}
\vec{\tau}\bar{D}+D\vec{\tau}\bar{D}^{*}_{\mu})\cdot\vec{\pi},\\  
{\cal L}_{\rho \psi DD} & = & g_{\rho \psi DD}\psi^{\mu}D\vec{\tau}\bar{D}
\cdot\vec{\rho}_{\mu}, \\
{\cal L}_{\rho \psi D^{*}D^{*}} & = & g_{\rho \Psi
D^{*}D^{*}}(\psi^{\nu}D^{*}_{\nu}\vec{\tau}\bar{D}^{*}_{\mu}+
\psi^{\nu}D^{*}_{\mu}\vec{\tau}\bar{D}^{*}_{\nu} \nonumber\\
&&-2\psi_{\mu}D^{*\nu}\vec{\tau}\bar{D}^{*}_{\nu})\cdot\vec{\rho}^{\mu}.  
\end{eqnarray}
In the above, ${\vec\tau}$ are Pauli spin matrices, and $\pi$ and
$\rho$, denote the pion and rho meson isospin triplet, respectively,
while $D=(D^0,D^+)$ and $D^*=({D^*}^0,{D^*}^+)$ denote the pseudoscalar
and vector charm meson doublets, respectively. The $J/\psi$ is denoted
by $\psi$ while $N$ represents the nucleon.

For coupling constants, we use the empirical values $g_{\pi NN}=13.5$
\cite{pnn}, $g_{\rho NN}=3.25$, and $\kappa_\rho=6.1$ \cite{rnn},
and $g_{\pi DD^*}=4.4$ \cite{pdd}. From the vector dominance model,
we have $g_{\rho DD}=g_{\rho D^*D^*}=2.52$ and 
$g_{\psi DD}=g_{\psi D^*D^*}=7.64$ \cite{lin}. For the four-point
coupling constants, we relate their values to the three-point coupling
constants using the SU(4) relations \cite{lin}, i.e.,
\begin{eqnarray}
g_{\pi \psi DD^*}= g_{\pi DD^*} g_{\psi DD}, ~~
g_{\rho \psi D D}= 2~g_{\rho DD} g_{\psi DD},\nonumber\\
g_{\rho \psi D^* D^*}=g_{\rho D^* D^*} g_{\psi D^* D^*}.
\end{eqnarray}

The amplitudes for the first two processes in Fig. \ref{diagram1} are 
given by 
\begin{eqnarray} 
M_1 & = & -ig_{\pi NN}\bar{N}(p_{3})\gamma_{5}N(p_{1})
\frac{1}{t-m^{2}_{\pi}}\nonumber\\
&\times&(M_{1a}+M_{1b}+M_{1c}),\\ 
M_{2} & = & g_{\rho NN}\bar{N}(p_{3})\left[\gamma^{\mu}
+i\frac{\kappa_{\rho}}{2m_{N}}\sigma^{\alpha\mu}
(p_{1}-p_{3})_{\alpha}\right]\nonumber\\
&\times& N(p_{1})\left[-g_{\mu\nu}+\frac{(p_1-p_3)_{\mu}
(p_1-p_3)_{\nu}}{m^{2}_{\rho}}\right]\nonumber\\
&\times&\frac{1}{t-m^{2}_{\rho}}
(M^{\nu}_{2a}+M^{\nu}_{2b}+M^{\nu}_{2c}),
\end{eqnarray}  
where $p_1$ and $p_3$ are the four momenta of the initial
and final nucleons, respectively. In the above, 
$M_{1a},M_{1b},M_{1c}$ are the amplitudes for the subprocess
$\pi \psi \to D^{*}\bar{D}$ in the top three diagrams 
of Fig. \ref{diagram1}, while $M^{\nu}_{2a},M^\nu_{2b},M^\nu_{2c}$ are 
the amplitudes for the subprocesses $\rho \psi \to D\bar{D}$ in the 
middle three diagrams. The amplitude for the third process has a similar
expression as that for the second process with 
$M^{\nu}_{2a},M^\nu_{2b},M^\nu_{2c}$ replaced by 
$M^{\nu}_{3a},M^\nu_{3b},M^\nu_{3c}$, which are the amplitudes
for the subprocess $\rho \Psi \to D^{*}\bar D^{*}$ in the bottom
three diagrams. Expressions for these amplitudes can be found 
in Ref. \cite{lin}.

The cross sections for these processes with three particles
in the final state can be expressed in terms of the off-shell
cross sections of the subprocesses described by the amplitudes 
$M_1$, $M_2$, and $M_3$. Following the method of Ref. \cite{yao} 
for the reaction $NN\to N\Lambda K$, the spin and isospin averaged 
differential cross sections for the first two processes in Fig. 
\ref{diagram1} can be written as 
\begin{eqnarray}
\frac{d\sigma_{\psi N\to ND^*\bar D}}{dtds_1} & = & \frac{g^{2}_{\pi NN}}
{16\pi^{2}sp^{2}_{i}}k\sqrt{s_{1}}(-t)\frac{F_{\pi NN}^{2}(t)}
{(t-m^{2}_{\pi})^{2}}\nonumber\\
&\times&\sigma_{\pi\psi\to D^*\bar D}(s_{1},t),\\ 
\frac{d\sigma_{\psi N\to ND\bar D}}{dtds_{1}} 
& = & \frac{3g^{2}_{\rho NN}}{32\pi^{2}sp^{2}_{i}}k\sqrt{s_{1}}
\frac{F_{\rho NN}^{2}(t)}{(t-m^{2}_{\rho})^{2}}
\left[4(1+\kappa_{\rho})^2\right .\nonumber\\
&\times&(-t-2m^{2}_{N})\kappa^{2}_{\rho}\frac{(4m^{2}_{N}-t)^{2}}{2m^{2}_{N}}
+4(1+\kappa_{\rho})\nonumber\\
&\times&\left .\kappa_{\rho}(4m^{2}_{N}-t)\right]
\sigma_{\rho\psi\to D\bar D}(s_{1},t),
\end{eqnarray}
and the differential cross section for $J\psi N\to D^*\bar D^*N$ is 
similar to that for $J\psi N\to D\bar DN$ with 
$\sigma_{\rho\psi\to D\bar D}(s_{1},t)$ replaced by 
$\sigma_{\rho\psi\to D^*\bar D^*}(s_{1},t)$.

In the above, $p_i$ is the center-of-mass momentum of $J/\psi$ and
$N$, $t$ is the squared four momentum transfer, and 
$s_{1}$ and $k$ are, respectively, the squared invariant 
mass and center-of-mass momentum of $\pi$ and $J/\psi$ in the process 
$J/\psi N\to D^*\bar DN$ or $\rho$ and $J/\psi$ in the processes 
$J/\psi N\to D\bar DN$ and $J/\psi N\to D^*\bar D^*N$. We have also
introduced form factors $F_{\pi NN}$ and $F_{\rho NN}$ at the $\pi NN$ 
and $\rho NN$ vertices, respectively. As in Ref.\cite{yao},
both are taken to have the monopole form, i.e., 
\begin{eqnarray}\label{form1}
F_1(t)=\frac{\Lambda^2-m^2}{\Lambda^2-t},
\end{eqnarray}
where $m$ is the mass of exchanged pion or rho meson, and 
$\Lambda$ is a cutoff parameter.  Following Refs.\cite{pnn,rnn},
we take $\Lambda_{\pi NN}=1.3$ GeV and $\Lambda_{\rho NN}=1.4$ GeV.

The cross sections $\sigma_{\pi\psi\to D^*\bar D}(s_{1},t)$,
$\sigma_{\rho\psi\to D\bar D}(s_{1},t)$, and
$\sigma_{\rho\psi\to D^*\bar D^*}(s_{1},t)$ are the 
spin and isospin averaged differential 
cross sections for the subprocesses $\pi\psi\to D^*\bar D$,
$\rho\psi\to D\bar D$, and $\rho\Psi\to D^*\bar D^*$ with off-shell
pion or rho meson. Explicit expressions for these cross sections
can be obtained from Ref. \cite{lin} by replacing the square of 
pion or rho meson masses by $t$. In evaluating these cross sections,
we also introduce form factors at the interaction vertices.
Following Ref.\cite{lin}, the form factors at   
three-point $t$ channel and $u$ channel vertices, i.e., 
$\pi DD^*$, $\rho DD$, $\rho D^*D^*$, $\psi DD$, and $\psi D^*D^*$
that involve heavy virtual charm mesons, are taken to have the following form:
\begin{eqnarray}\label{form2}
F_2({\bf q}^2)=\frac{\Lambda^2}{\Lambda^2+{\bf q}^2},
\end{eqnarray}
instead of the monopole form of Eq. (\ref{form1}).
In the above, ${\bf q}$ is the three momentum transfer in the
center-of-mass of $\psi$ and pion or rho meson.

The form factor at four-point vertices, i.e., $\pi\psi DD^*$, $\rho\psi DD$,
and $\rho\psi D^*D^*$, are taken to be
\begin{eqnarray}
f_4=\left ( \frac {\Lambda_1^2}{\Lambda_1^2+<{{\bf q}^2}> } \right )
\left ( \frac {\Lambda_2^2}{\Lambda_2^2+<{{\bf q}^2}> } \right ),
\end{eqnarray}
where $\Lambda_1$ and $\Lambda_2$ are
the two different cutoff parameters at the three-point vertices 
present in processes with the same initial and final particles, 
and $<{{\bf q}^2}>$ is the average value 
of the squared three momentum transfers in $t$ and $u$ channels.

\begin{figure}[h]
\centerline{\epsfig{file=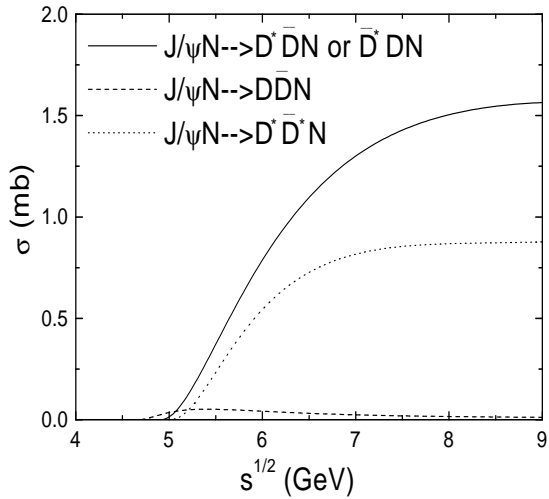,width=3in,height=3.2in,angle=-90}}
\vspace{0.5cm}
\caption{$J/\psi$ absorption cross sections by nucleon due to the 
virtual pion and rho meson cloud of the nucleon as functions
of center-of-mass energy.}
\label{cross1}
\end{figure}

Using the same value of 1 GeV for cutoff parameters 
in the form factors involving charm mesons as in Refs. 
\cite{lin,oh}, we have evaluated the cross sections for
$J/\psi$ absorption by nucleon, and they are shown in 
Fig. \ref{cross1} as functions of total center-of-mass energy.
It is seen that all cross sections are less than 2 mb. 
Furthermore, the cross section for $J/\psi N\to D^*\bar DN$
and $J/\psi N\to \bar D^*DN$
(solid curve) due to pion exchange is larger than
those for $J/\psi N\to D\bar DN$ (dashed curve) and 
$J/\psi N\to D^*\bar D^*N$ (dotted curve) that are due to rho meson exchange. 

Our result for $\sigma_{J/\psi N\to D\bar DN}$ is
order-of-magnitude smaller than that of Ref. \cite{sib}, where this processes 
is viewed as the elastic scattering of a nucleon with one of
the charm mesons from the decay of $J/\psi$. The latter cross
section is then assumed to have a constant value of 20 mb. 
Compared to our approach, they have neglected both 
the energy dependence and the off-shell effect of the 
subprocess involved in $J/\psi-N$ absorption to three-body
final state.  Also contributing to this large difference in
the cross section is the value of cutoff parameter, 3.1 GeV in 
Ref.\cite{sib} versus 1 GeV used here, and the different momentum
dependence, four momentum transfer in Ref.\cite{sib} while 
three momentum transfer in the present study. We note that
the more important processes $J/\Psi\to D^*\bar DN(\bar D^*DN)$
and $J/\Psi\to D^*\bar D^*N$ are not considered in Ref.\cite{sib}.

\section{$J/\psi$ absorption by nucleon via charm exchange}\label{charm}

\begin{figure}[h]
\centerline{\epsfig{file=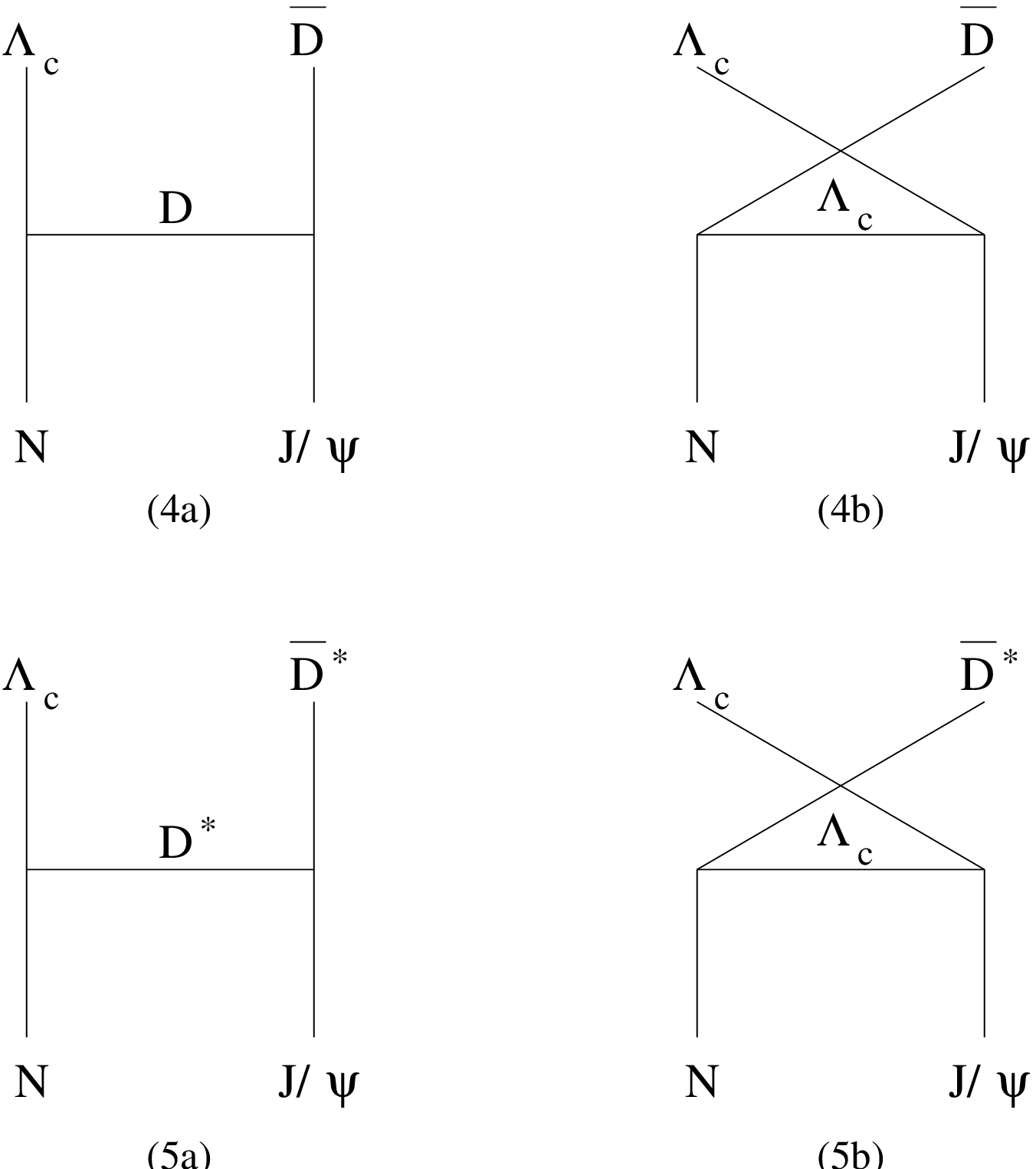,width=2in,height=2.5in,angle=0}}
\vspace{0.5cm}
\caption{$J/\psi$ absorption by nucleon via charm exchange.}
\label{diagram2}
\end{figure}

Besides absorption by the virtual pion and rho meson cloud of a nucleon,
$J/\psi$ can also be absorbed by the nucleon via charm exchange
in the reaction $J\psi N\to\bar D\Lambda_c$ and
$J\psi N\to\bar D^*\Lambda_c$ shown by
the diagrams in Fig. \ref{diagram2}. These processes involve 
the following interaction Lagrangians: 
\begin{eqnarray}
{\cal L}_{DN\Lambda_{c}} & = &ig_{DN\Lambda_{c}}(\bar{N}\gamma_{5}\Lambda_{c}\bar D+D
\bar{\Lambda}_{c}\gamma_{5}N),\label{dnl}\\
{\cal L}_{D^{*}N\Lambda_{c}} & = &
g_{D^{*}N\Lambda_{c}}(\bar{N}\gamma_{\mu}\Lambda_{c}\bar D^{*\mu}
+D^{*\mu}\bar{\Lambda}_{c} \gamma_{\mu}N),\label{dsnl}\\
{\cal L}_{\Psi\Lambda_{c}\Lambda_{c}} & = & g_{\psi\Lambda_{c}\Lambda_{c}}
\bar{\Lambda}_{c}\gamma^{\mu}\psi_{\mu}\Lambda_{c}.\label{jll}
\end{eqnarray}
where $\Lambda_c$ denotes the charm baryon.

The amplitudes for these processes are given by 
\begin{eqnarray} 
M_{4a}&=&M^{\mu}_{4a}\varepsilon_{2\mu},\\
M_{4b}&=&M^{\mu}_{4b}\varepsilon_{2\mu},\\
M_{5a}&=&M^{\mu\nu}_{5a}\varepsilon_{2\mu}\varepsilon_{4\nu}; \\ 
M_{5b}&=&M^{\mu\nu}_{5b}\varepsilon_{2\mu}\varepsilon_{4\nu}. 
\end{eqnarray}
with $\varepsilon_{2\mu}$ and $\varepsilon_{4\mu}$
being the polarization vectors of $J/\psi$ and $D^*$, respectively, and
\begin{eqnarray}
M^{\mu}_{4a} & = &  2ig_{\psi DD}g_{DN\Lambda_{c}}\frac{1}{t-m^{2}_{D}}
p^{\mu}_{4}\bar{\Lambda}_{c}(p_{3})\gamma_{5}N(p_{1}),\\    
M^{\mu}_{4b} & = & ig_{DN\Lambda_{c}}g_{\psi\Lambda_{c}\Lambda_{c}}
\bar{\Lambda}_{c}(p_{3})\gamma^{\mu}\frac{{q\mkern-10mu/}
+m_{\Lambda_{c}}}{u-m^{2}_{\Lambda_{c}}}\gamma_{5}N(p_{1}),\nonumber\\ 
M^{\mu\nu}_{5a} & = & -g_{D^{*}N\Lambda_{c}}g_{\psi D^{*}D^{*}}
\bar{\Lambda}_{c}(p_{3})\gamma^{\alpha}N(p_1)\nonumber\\
&\times&\left[g_{\alpha\beta}-\frac{(p_{1}-p_{3})_{\alpha}
(p_{1}-p_{3})_{\beta}}{m^{2}_{D^{*}}}\right]\nonumber\\   
&\times&\frac{1}{t-m^{2}_{D^{*}}}[p^{\nu}_{2}g^{\beta\mu}
-(p_{2}+p_{4})^{\beta}g^{\mu\nu}+2p^{\mu}_{4}g^{\beta\nu}],\\ 
M^{\mu\nu}_{5b} & = &g_{D^{*}N\Lambda_{c}}g_{\psi\Lambda_{c}
\Lambda_{c}}\bar{\Lambda}_{c}(p_{3})\gamma^{\mu}\frac{{q\mkern-10mu/}
+m_{\Lambda_{c}}}{u-m^{2}_{\Lambda_{c}}}\gamma^{\nu}N(p_{1}).  
\end{eqnarray}
In the above, $q=p_1-p_4$, and $s=(p_1+p_2)^2$ and $t=(p_1-p_3)^2$ are 
the standard Mendelstam variables.

The spin and isospin averaged differential cross sections for 
these two-body processes are then
\begin{eqnarray}
\frac{d\sigma_{\psi N\rightarrow\bar{D}\Lambda_{c}}}{dt} 
& = & \frac{1}{64\pi sp^{2}_{i}}|M_{4a}+M_{4b}|^2,\label{diff1}\\
\frac{d\sigma_{\psi N\rightarrow\bar{D}^{*}\Lambda_{c}}}{dt}
& = &\frac{1}{64\pi sp^{2}_{i}}|M_{5a}+M_{5b}|^2,\label{diff2}
\end{eqnarray} 
where $|M_{4a}+M_{4b}|^2$ and $|M_{5a}+M_{5b}|^2$
can be evaluated using the software package FORM \cite{form}. 

\begin{figure}[h]
\centerline{\epsfig{file=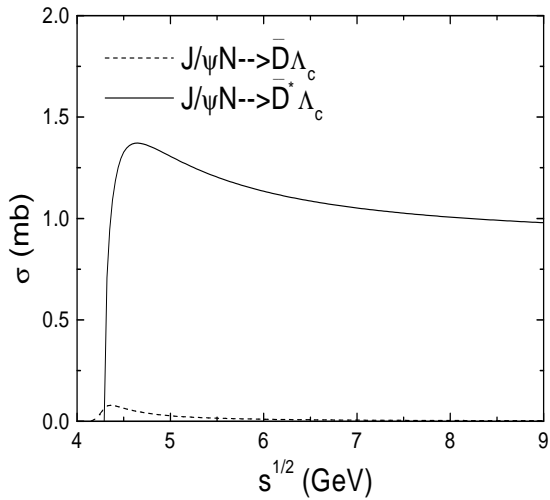,width=3in,height=3.2in,angle=-90}}
\vspace{0.5cm}
\caption{$J/\psi$ absorption cross sections by nucleon 
due to charm exchange as functions of center-of-mass energy.}
\label{cross2}
\end{figure}

The coupling constants $g_{DN\Lambda_c }$, $g_{D^*N\Lambda_c}$,
and $g_{\psi\Lambda_c\Lambda_c}$ can be related to known
coupling constants $g_{\pi NN}$ and $g_{\rho NN}$
using the SU(4) symmetry as shown in the Appendix. 
Using $g_{\pi NN}=13.5$ and $g_{\rho NN}=3.25$, 
we then have $g_{DN\Lambda_c}=13.5$, $g_{D^*N\Lambda_c}=-5.6$,
and $g_{\psi\Lambda_c\Lambda_c}=-1.4$. 
We again introduce monopole form factors of Eq. (\ref{form2}) at the 
vertices with cutoff parameter $\Lambda=1$ GeV.
The resulting cross sections for $\psi N\to\bar D\Lambda_c$ 
and $\psi N\to\bar D^*\Lambda_c$ are shown in Fig. \ref{cross2} 
by the dashed and solid curves, respectively. 
Their values are seen to be less than 1 mb. Furthermore, 
$\sigma_{J/\psi N\to\bar D^*\Lambda_c}$ is much larger than
$\sigma_{J/\psi N\to\bar D\Lambda_c}$ due to the three vector
mesons coupling, which has been shown to increase significantly 
the $J/\psi-\pi$ absorption cross section as well \cite{lin}.

In Ref. \cite{sib}, only diagram (4a) in Fig. \ref{diagram2} 
has been studied, and the result there is about a factor of 
4 larger than our cross section for $J/\psi N\to \bar D\Lambda_c$,
which includes also diagram (4b). The larger cross section in Ref.
\cite{sib} is again due to both a larger cutoff parameter of 2 GeV
versus 1 GeV used here and the use of four momentum 
instead of three momentum transfer in the form factors. Our total $J/\psi-N$ 
absorption cross section due to charm exchange is, however, 
larger as we have also included the more important processes 
shown by diagrams (5a) and 5(b).

\section{anomalous parity interactions}\label{anomalous}

\begin{figure}[h]
\centerline{\epsfig{file=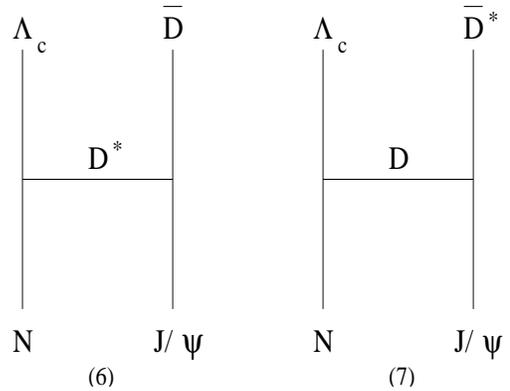,width=2.6in,height=2in,angle=0}}
\vspace{0.5cm}
\caption{$J/\psi$ absorption cross section by nucleon via charm meson
exchange through the anomalous parity interaction.}
\label{diagram3}
\end{figure}

There are also anomalous parity interactions of $J/\psi$ with charm
mesons \cite{oh}, i.e.,
\begin{eqnarray}
{\cal L}_{\psi D^{*}D} & = & g_{\psi
D^{*}D}\varepsilon_{\alpha\beta\mu\nu}(\partial^{\alpha}\psi^{\beta})
[(\partial^{\mu}\bar{D}^{*\nu})D \nonumber\\
&+&\bar{D}(\partial^{\mu}D^{*\nu})],
\end{eqnarray}
which not only introduces additional diagrams for the processes
shown in Fig. \ref{diagram1} but also leads to the reactions 
$J/\psi N\to\bar D\Lambda_c$ via $D^*$ exchange and 
$J/\psi N\to\bar D^*\Lambda_c$ via $D$ exchange shown by the 
diagrams in Fig. \ref{diagram3}. 

As shown in Ref. \cite{oh}, the anomalous interaction is not
important for $J/\psi-\rho$ absorption and increases the 
$J/\psi-\pi$ absorption cross section by only about 50\%.
Thus, inclusions of additional diagrams due to the anomalous parity 
interactions in processes involving three-body final states shown
in Fig. \ref{diagram1} will probably increase the $J/\psi-N$ absorption 
cross section calculated here by less than 50\%. 

The amplitudes for the process $J/\psi N\to\bar D\Lambda_c$ and
$J/\psi N\to\bar D^*\Lambda_c$ are given by 
\begin{eqnarray}
M_6 &=& M^{\mu}_6\varepsilon_{2\mu}, \\
M_7 &=& M^{\mu\nu}_7\varepsilon_{2\mu}\varepsilon_{4\nu},
\end{eqnarray} 
with $\varepsilon_{2\mu}$ and $\varepsilon_{4\nu}$ being the
polarization vectors of $J/\psi$ and $D^*$, respectively, and 
\begin{eqnarray} 
M^{\mu}_{6} & = & -g_{\psi D^{*}D}g_{D^{*}N\Lambda_{c}}
\frac{1}{t-m^{2}_{D^{*}}}\varepsilon^{\mu\nu\alpha\beta}
p_{2\alpha}(p_{1}-p_{3})_{\beta}\nonumber\\
&\times&\bar{\Lambda}_{c}(p_{3})\gamma_{\nu}N(p_{1}), \\
M^{\mu\nu}_{7} & = & ig_{\psi D^{*}D}g_{DN\Lambda_{c}}
\frac{1}{t-m^{2}_{D}}\varepsilon^{\mu\nu\alpha\beta}
p_{2\alpha}p_{4\beta}\nonumber\\
&\times&\bar{\Lambda}_{c}(p_{3})\gamma_{5}N(p_{1}).  
\end{eqnarray} 

Because of the anomalous parity in the $\Psi D^*D$ vertex, the process 
$J/\Psi N\to\bar D\Lambda_c$ via $D^*$ exchange 
does not interfere with the similar process
via $D$ exchange shown in Fig. \ref{diagram2}. 
The differential cross sections for the two anomalous
processes in Fig. \ref{diagram3} are given by similar expressions 
as Eqs. (\ref{diff1}) and (\ref{diff2}) with

\begin{eqnarray}
|M_{6}|^{2} & = & \frac{g^{2}_{\psi D^{*}D}g^{2}_{D^{*}N\Lambda_{c}}}
{12m^{2}_{\psi}}\frac{1}{(t-m^{2}_{D^{*}})^{2}}
\left\{4m^{2}_{\psi}[2(m^{2}_{N}\right.\nonumber\\
&+&m^{2}_{\Lambda_{c}})t-t^{2}-(m^{2}_{\Lambda_{c}}-m^{2}_{N})^{2}]
+2(m^{2}_{\Lambda_{c}}-m^{2}_{N})\nonumber\\
&\times&[(m^{2}_{\psi}+m^{2}_{\Lambda_{c}}-u)^{2}
-(s-m^{2}_{N}-m^{2}_{\psi})^{2}]\nonumber\\
&-&[2(m^{2}_{N}+m^{2}_{\Lambda_{c}})-t](2m^{2}_{\psi}
+m^{2}_{\Lambda_{c}}+m^{2}_{N}-u-s)^{2}\nonumber\\
&-&t(m^{2}_{\Lambda_{c}}-m^{2}_{N}+s-u)^{2}
-2[(m_{N}-m_{\Lambda_{c}})^{2}-t]\nonumber\\
&\times&\left.[4m^{2}_{\psi}t-(2m^{2}_{\psi}+m^{2}_{N}
+m^{2}_{\Lambda_{c}}-u-s)^{2}]\right\},
\end{eqnarray}
and
\begin{eqnarray}
|M_{7}|^{2} & = & \frac{g^{2}_{\Psi D^{*}D}g^{2}_{DN\Lambda_{c}}}
{6m^{2}_{\psi}}\frac{1}{(t-m^{2}_{D})^{2}}
[(m_{N}-m_{\Lambda_{c}})^{2}-t]\nonumber\\
&\times&[(m^{2}_{\psi}+M^{2}_{D^{*}}-t)^{2}-m^{2}_{\psi}m^{2}_{D^{*}}],
\end{eqnarray}
where $u=(p_1-p_4)^2$.

\begin{figure}[h]
\centerline{\epsfig{file=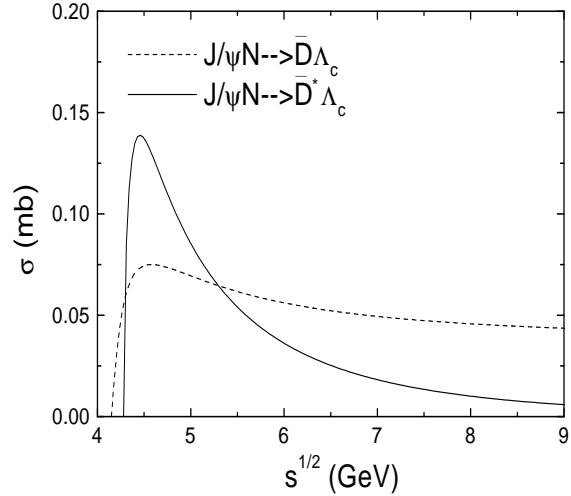,width=3in,height=3.2in,angle=-90}}
\vspace{0.5cm}
\caption{Contributions of the anomalous interaction to 
$J/\psi$ absorption cross sections by nucleon as functions
of center-of-mass energy.}
\label{cross3}
\end{figure}

The coupling constant 
in the anomalous parity interaction has been determined 
to be $g_{\psi DD^*}=8.61$ GeV$^{-1}$ from the 
radiative decay of $D^*$ to $D$ using the vector dominance model
\cite{oh}. With a monopole form factor similar to 
Eq. (\ref{form2}) at the $D^*N\Lambda_c$ vertex and a cutoff 
parameter of 1 GeV, the cross sections for the reactions
$J/\psi N\to D\Lambda_c$ due to $D^*$ exchange and 
$J/\psi N\to D^*\Lambda_c$ due to $D$ exchange
are shown in Fig. \ref{cross3}.  Their values are seen to be less
than 0.15 mb, which is negligible compared to the contributions from 
the normal interactions studied in Sections \ref{pion} and \ref{charm}.

We note that the two processes in Fig. \ref{diagram3} due to the
anomalous interaction have also been studied in Ref. \cite{sib}.
Their coupling constant is related to ours by $g_{\psi DD^*}/m_{J/\psi}$,
where $m_{J/\psi}$ is the mass of $J/\psi$. Since they assume that
$g_{\psi DD^*}=g_{\psi DD}=7.64$ based on an incorrect quotation 
from Ref. \cite{lin1}, the strength of the anomalous coupling constant
in their study is only 2.47 GeV$^{-1}$ and is about a factor
of 3 smaller than that used here. However, they have used a
much larger value for $g_{D^*N\Lambda_c}=-19$ than that
given by the SU(4) relation. As a result, 
their cross section for diagram (7) in Fig. \ref{diagram3} 
should have a similar magnitude as ours while that of diagram (6)
should be larger than our value. Because of the larger value of
cutoff parameter of 2 GeV and the use of four momentum transfer in the
form factor, the results in Ref.\cite{sib} from the anomalous interaction
turn out to be order-of-magnitude larger than ours.

\section{total $J/\psi$ absorption cross section by nucleon}\label{total}

\begin{figure}[h]
\centerline{\epsfig{file=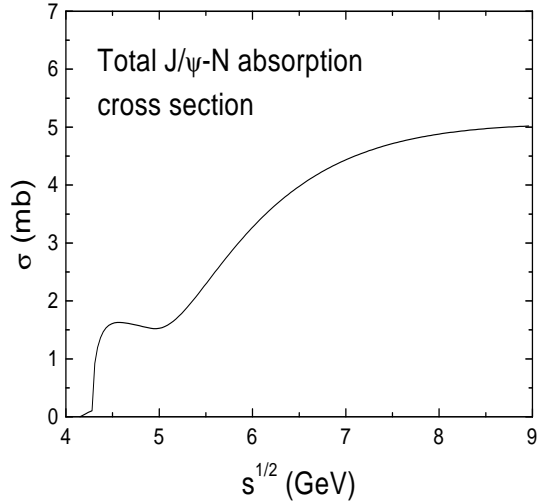,width=3in,height=3.2in,angle=-90}}
\vspace{0.5cm}
\caption{Total $J/\psi$ absorption cross sections by nucleon as functions
of center-of-mass energy.}
\label{cross4}
\end{figure}

The total $J/\psi$ absorption cross section by nucleon, obtained
by adding the contributions shown in Figs. \ref{cross1} and \ref{cross3}
is given in Fig. \ref{cross4}. At low center-of-mass energies, the
cross section is dominated by the process $J/\psi N\to\bar D^*\Lambda_c$
while at high center-of-mass energies, the processes $J/\psi N\to D^*\bar DN$
and $J/\psi N\to \bar D^*DN$ due to the virtual pion cloud of the nucleon
are most important. The total $J/\psi$ absorption cross section
is at most 5 mb and is consistent with that extracted from $J/\psi$ 
production in photo-nucleus and proton-nucleus reactions.

\section{conclusions and discussions}\label{conclusion}

We have used a meson-exchange model to study the $J/\psi$ absorption
cross section by a nucleon. The interaction Lagrangians are based
on the gauged SU(4) flavor symmetry but with empirical masses. 
Using coupling constants taken either from the empirical information
or via the SU(4) relations and form factors with cutoff parameter 
of 1 GeV, we obtain a $J/\psi$-nucleon absorption 
cross section of at most 5 mb, which is consistent
with the empirical cross section extracted from $J/\psi$ production
in photo-nucleus and proton-nucleus reactions.
Since the dominant process can be viewed
as $J/\psi$ absorption by the virtual pion cloud
of a nucleon, our results thus indicate that the cross sections for $J/\psi$ 
absorption by pion and rho meson evaluated in previous 
studies using the meson-exchange model are not in contradiction with 
the empirical cross section for $J/\psi$ absorption by nucleon. 

Our results are not much affected if we use the coupling constants
$g_{DN\Lambda_c}\sim 6.7-7.9$ and $g_{D^*N\Lambda_c}\sim -7.5$
determined from the QCD sum rules \cite{qcd} instead from the
SU(4) symmetry. With these values,
$\sigma_{\psi N\to\bar D\Lambda_c}$ will be even smaller 
while $\sigma_{\psi N\to\bar D^*\Lambda_c}$ will be about a factor of 
two larger than those shown in Fig. \ref{cross2}. In this case,
the $J/\psi-N$ absorption cross section is only increased by 
about 1 mb. On the other hand, if the cutoff parameter is taken 
to be $\Lambda=2$ GeV at vertices involving charm hadrons as suggested 
by QCD sum rules \cite{qcd}, then the total $J/\psi-N$ absorption cross 
section increases to about 10 mb, which is about a factor of two larger 
than the empirical value from $J/\psi$ production in 
photo-nucleus and proton-nucleus reactions. With this cutoff
parameter, the $J/\psi-\pi$ absorption cross section is also about 10 mb
as shown in Ref.\cite{lin}. Since the meson-exchange model is
based on effective hadronic Lagrangians, one can either 
fit the empirical $J/\psi-N$ absorption cross section by
treating the cutoff parameter as a phenomenological parameter,
or use the cutoff parameter from the QCD sum rules but with a
different effective Lagrangian. In the former case, a 
cutoff parameter of 1 GeV is required at the interaction vertices 
involving charm hadrons in order to have the correct $J/\psi_N$ 
absorption cross section. The meson-exchange model of
Ref.\cite{lin} then gives a $J/\psi-\pi$ absorption cross section
of about 3 mb, which is also consistent with that used in the comover model
for $J/\psi$ suppression in heavy ion collisions \cite{comover,cassing}.
In the latter case, one may follow the suggestion of Ref.\cite{chiral}
to drop the nongradient pion couplings in the effective Lagrangians,
as they  breaks the chiral $SU(2)\times SU(2)$ symmetry.
As shown in Ref. \cite{chiral}, neglecting these terms reduces 
the $J/\psi-\pi$ absorption cross section by about a factor of two,
leading again to a $J/\psi-\pi$ absorption cross section
similar to that in the comover model. The $J/\psi-N$ absorption cross 
section obtained with such an effective Lagrangian is expected to be 
reduced as well.
  
\section*{acknowledgment}

This work was supported by the National Science Foundation under Grant 
No. PHY-9870038, the Welch Foundation under Grant No. A-1358, and the 
Texas Advanced Research Program under Grant No. FY99-010366-0081. 

\section*{appendix}

In the SU(4) quark model, baryons belong to the 20-plet states.
These states can be conveniently expressed by tensors 
$\phi_{\mu\nu\lambda}$ \cite{okubo}, where $\mu$, $\nu$, and 
$\lambda$ run from 1 to 4, that satisfy the conditions
\begin{eqnarray}
\phi_{\mu\nu\lambda}+\phi_{\nu\lambda\mu}+\phi_{\lambda\mu\nu}=0,\quad
\phi_{\mu\nu\lambda}=\phi_{\nu\mu\lambda}.
\end{eqnarray}

For baryons without charm quarks, i.e., belonging to SU(3) octet,
they are given by
\begin{eqnarray}
p&=&\phi_{112},\quad n=\phi_{221},\quad
\Lambda=\sqrt{\frac{2}{3}}(\phi_{321}-\phi_{312}),\nonumber\\
\Sigma^+&=&\phi_{113},\quad\Sigma^0=\sqrt{2}\phi_{123}\quad\Sigma^-=\phi_{223},
\nonumber\\
\quad\Xi^0&=&\phi_{331},\quad\Xi^-=\phi_{332}.
\end{eqnarray}

For baryons with one charm quark, they are
\begin{eqnarray}
\Sigma_c^{++}&=&\phi_{114},\quad\Sigma_c^+=\phi_{124},
\quad\Sigma_c^0=\phi_{224},\nonumber\\
\Xi_c^+&=&\phi_{134},\quad\Xi_c^0=\phi_{234},\nonumber\\
\Xi_c^{+\prime}&=&\sqrt{\frac{2}{3}}(\phi_{413}-\phi_{431}),\quad
\Xi_c^{0\prime}=\sqrt{\frac{2}{3}}(\phi_{423}-\phi_{432}),\nonumber\\
\Lambda_c^+&=&\sqrt{\frac{2}{3}}(\phi_{421}-\phi_{412}),
\quad\Omega_c^0=\phi_{334}.
\end{eqnarray}

For baryons with two charm quarks, they are
\begin{eqnarray}
\Xi_{cc}^{++}&=&\phi_{441},\quad\Xi_{cc}^+=\phi_{442},\quad
\Omega_{cc}^+=\phi_{443}.
\end{eqnarray}

Mesons in the SU(4) quark model belong to the 15-plet.
In the tensor notations, pseudoscalar and vector mesons are
expressed by $P^{\alpha}_{\beta}$ and $V^{\alpha}_{\beta}$, respectively.
For pseudoscalar mesons, we have
\begin{eqnarray}
\pi^+&=&P^2_1,\quad\pi^-=P^1_2,
\quad\pi^0=\frac{1}{\sqrt{2}}(P^1_1-P^2_2),\nonumber\\
K^+&=&P^3_1,\quad K^0=P^3_2,\quad K^-=P^1_3,\quad{\bar K^0}=P^2_3,\nonumber\\
D^+&=&P^2_4,\quad D^0=P^1_4,\quad D^-=P^4_2,\quad{\bar D^0}=P^4_1,\nonumber\\
D_s^+&=&P^3_4,\quad D_s^-=P^4_3,\nonumber\\
\eta&=&\frac{1}{\sqrt{6}}(P^1_1+P^2_2-2P^3_3),\nonumber\\
\eta_c&=&\frac{1}{\sqrt{12}}(P^1_1+P^2_2+P^3_3-3P^4_4).
\end{eqnarray} 

Similarly, we have for vector mesons
\begin{eqnarray}
\rho^+&=&V^2_1,\quad\rho^-=V^1_2,\quad
\rho^0=\frac{1}{\sqrt{2}}(V^1_1-V^2_2),\nonumber\\
K^{*+}&=&V^3_1,\quad K^{*0}=V^3_2,\quad K^{*-}=V^1_3,
\quad{\bar K^{*0}}=V^2_3,\nonumber\\
D^{*+}&=&V^2_4,\quad D^{*0}=V^1_4,\quad D^{*-}=V^4_2,
\quad \bar D^{*0}=V^4_1,\nonumber\\
D_s^{*+}&=&V^3_4,\quad D_s^{*-}=V^4_3,\nonumber\\
\omega&=&\frac{1}{\sqrt{6}}(V^1_1+V^2_2-2V^3_3),\nonumber\\
J/\psi&=&\frac{1}{\sqrt{12}}(V^1_1+V^2_2+V^3_3-3V^4_4).
\end{eqnarray}

In tensor notations, the SU(4) invariant interaction Lagrangians 
between baryons and pseudoscalar mesons as well as between baryons 
and vector mesons can be written, respectively, as 
\begin{eqnarray}
{\cal L}_{PBB}&=&g_p(a\phi^{*\alpha\mu\nu}\gamma_{5}P^{\beta}_{\alpha}
\phi_{\beta\mu\nu}\nonumber\\ 
&&+b\psi^{*\alpha\mu\nu}\gamma_{5}P^{\beta}_{\alpha}\phi_{\beta\nu\mu}),\\
{\cal L}_{VBB}&=&g_v(c\phi^{*\alpha\mu\nu}\gamma\cdot
V^{\beta}_{\alpha}\phi_{\beta\mu\nu}\nonumber\\ 
&&+d\phi^{*\alpha\mu\nu}\gamma\cdot
V^{\beta}_{\alpha}\phi_{\beta\nu\mu}),
\end{eqnarray}
where $g_p$ and $g_v$ are the universal baryon-pseudoscalar-meson and
baryon-vector-meson coupling constants, and $a$, $b$, $c$, and $d$ 
are constants.

Writing explicitly, we obtain the following interaction Lagrangians, 
\begin{eqnarray}
{\cal L}_{PBB}&=&g_p\left[\frac{1}{\sqrt{2}}(a-\frac{5}{4}b)\bar N\gamma_5
\vec\tau\cdot\vec\pi N\right .\nonumber\\
&+&\left .\frac{3\sqrt{6}}{8}(b-a)(\bar{N}\gamma_{5}K\Lambda+
\bar N\gamma_{5}\bar D\Lambda_{c})+\cdots\right],\label{lp}\\
{\cal L}_{VBB}&=&g_v\left[\frac{1}{\sqrt{2}}(c-\frac{5}{4}d)\bar{N}
\gamma_\mu\rho^\mu N\right .\nonumber\\
&+&\frac{3\sqrt{6}}{8}(d-c)(\bar N\gamma_\mu K^{*\mu}\Lambda+
\bar{N}\gamma_{\mu}\bar D^{\mu*}\Lambda_{c})\nonumber\\
&+&\left .\frac{\sqrt{3}}{4}(-c+\frac{3}{2}d)\bar{\Lambda}_{c}
\gamma_{\mu}\psi^{\mu}\Lambda_{c}+\cdots\right].\label{lv}
\end{eqnarray}

The baryon-pseudoscalar-meson coupling in SU(3) is usually written
as $DTr[(B\bar B+\bar BB)M]+FTr[(B\bar B-\bar BB)M]$, where $B$ and $M$
are the baryon and pseudoscalar meson octet. In terms of the ratio
$\alpha_D=D/(D+F)$, which has an empirical value of about 0.64 \cite{alpha},
we then have the following relation between $g_{\pi NN}$ and
$g_{KN\Lambda}$ in the Lagrangians ${\cal L}_{\pi NN}$ given by Eq. (\ref{pnn})
and ${\cal L}_{KN\Lambda}=ig_{KN\Lambda}\bar N\gamma_5\Lambda\bar K$: 
\begin{eqnarray}
g_{KN\Lambda}=\frac{3-2\alpha_D}{\sqrt{3}}g_{\pi NN},
\end{eqnarray}
Comparisons with the SU(4) relations in Eq. (\ref{lp}) then gives
\begin{equation}\label{ab}
\frac{b}{a}=\frac{3-8\alpha_D}{6-10\alpha_D}.
\end{equation}

The baryon-vector-meson coupling is usually introduced through 
minimal coupling by treating vector mesons as gauge particles. 
In SU(3), this leads to the following relation between $g_{\rho NN}$ and
$g_{K^*N\Lambda}$ in the Lagrangians ${\cal L}_{\rho NN}$ given by 
Eq. (\ref{rnn}) and 
${\cal L}_{K^*N\Lambda}=g_{K^*N\Lambda}\bar N\gamma_\mu\Lambda\bar K^*$:
\begin{equation}
g_{K^{*}N\Lambda}=-\sqrt{3}g_{\rho NN}.
\end{equation}
Comparing with the SU(4) relations in Eq. (\ref{lv}), we have
\begin{equation}\label{cd}
\frac{d}{c}=\frac{1}{2}.
\end{equation}

Using Eqs. (\ref{ab}) and (\ref{cd}) in Eqs. (\ref{lp}) and (\ref{lv}),
we then have
\begin{eqnarray}
g_{DN\Lambda_c}&=&\frac{3-2\alpha_D}{\sqrt{3}}g_{\pi NN},\nonumber\\
g_{\psi\Lambda_{c}\Lambda_{c}}&=&-\frac{g_{\rho NN}}
{\sqrt{6}},\quad g_{D^{*}N\Lambda{c}}=-\sqrt{3}g_{\rho NN},
\end{eqnarray} 
for the coupling constants in the Lagrangians given by 
Eqs. (\ref{dnl}), (\ref{dsnl}), and (\ref{jll}).

\end{multicols}


\begin{thebibliography}{99} 
\bibitem{jpsi}M. Gonin {\it et al.} (NA50 Collaboration), Phys. Lett.
B 450, 456 (1999).
\bibitem{matsui}T. Matsui and H. Satz, Phys. Lett. B 178, 416 (1986).
\bibitem{comover}S. Gavin, Nucl. Phys. A 566, 287c (1994);
R. Vogt, Phys. Rep. 310, 197 (1999).
\bibitem{kharzeev}D. Kharzeev and H. Satz, Phys. Lett. B 334, 155 (1994).
\bibitem{qem}K. Martins, D. Blaschke, and E. Quack, Phys. Rev. C 51, 2723
(1995); C. Y. Wong, E. S. Swanson, and T. Barnes, Phys. Rev. C 62, 
045201 (2000).
\bibitem{haglin}K. Haglin, Phys. Rev. C 61, 031902 (2000).
\bibitem{lin}Z. W. Lin and C. M. Ko, Phys. Rev. C 62, 034903 (2000).
\bibitem{oh}Y. Oh, T. Song, and S. H. Lee, Phys. Rev. C 63, 034901 (2000).
\bibitem{nielsen}F. S. Navarra, M. Nielsen, R. S. Marques de Carvalho, and 
G. Krein, nucl-th/0105058.
\bibitem{photo}R. L. Anderson {\it et al.}, Phys. Rev. Lett. 38, 263 (1977).
\bibitem{pa}D. Kharzeev, C. Lourenco, M. Nardi, and H. Satz, Z. Phys.
C 74, 307 (1997).
\bibitem{lin2}Z. W. Lin, T. G. Di, and C. M. Ko, Nucl. Phys. A 689, 965 (2001).
\bibitem{pnn}B. Holzenkamp, K. Holinde, and J. Speth, Nucl. Phys. 
A 500 (1989) 485; G. Janssen, J.W. Durso, K. Holinde, B. C. Pearce, and
J. Speth, Phys. Rev. Lett. 71, 1978 (1993).
\bibitem{rnn}G. Janssen, K. Holinde, and J. Speth, Phys. Rev. C 54, 2218
(1996).
\bibitem{pdd}S. G. Matinyan and B. M\"uller, Phys. Rev. C 58, 2994 (1998).
\bibitem{yao}T. Yao, Phys. Rev. 125, 1048 (1961).
\bibitem{sib}A. Sibirtsev, K. Tsushima, and A. W. Thomas, Phys. Rev.
C 63, 044906 (2001).
\bibitem{form}J. Vermaseren, computer code FORM, 1989. Free version
of the software is available on the Internet at 
ftp://hep.itp.tuwien.ac.at/pub/FORM/PC/.
\bibitem{lin1}Z. W. Lin, C. M. Ko, and B. Zhang, Phys. Rev. C 61, 024904
(2000). 
\bibitem{qcd}F. S. Navarra and M. Nielsen, Phys. Lett. B 443, 285 (1998);
F. O. Dur\~aes, F. S. Navarra and M. Nielsen, {\it ibid.} 498, 169 (2001).
\bibitem{cassing}W. Cassing and C.M. Ko, Phys. Lett.B 396, 39 (1997); 
W. Cassing and E.L. Bratkovkaya, Nucl. Phys. A 623, 570 (1997). 
\bibitem{chiral}F. S. Navarra, M. Nielsen, and M. R. Robilotta, Phys. Rev.
C 64, 021901 (2001).
\bibitem{okubo}S. Okubo, Phys. Rev. D 11, 3261 (1975).
\bibitem{alpha}R. A. Adelseck and B. Saghai, Phys. Rev. C 42, 108 (1990).

\end{thebibliography}
\end{document}